# A Quest for the Structure of Intra- and Postoperative Surgical Team Networks: Does the Small World Property Evolve over Time?

Ashkan Ebadi[1], Patrick J. Tighe[2], Lei Zheng[2], and Parisa Rashidi[1]

[1]Department of Biomedical Engineering, University of Florida, Gainesville, FL, USA
[2]Department of Anesthesiology, University of Florida, Gainesville, FL, USA

## Abstract

**Objective:** We examined the structure of intra- and postoperative case-collaboration networks among the surgical service providers in a quaternary-care academic medical center, using retrospective electronic medical record (EMR) data. We also analyzed the evolution of the network properties over time, as changes in nodes and edges can affect the network structure.

**Materials and Methods:** We used de-identified intra- and postoperative data for adult patients, ages $\geq 21$, who received nonambulatory/nonobstetric surgery at Shands at the University of Florida between June 1, 2011 and November 1, 2014. The intraoperative segment contained 30,245 surgical cases, and the postoperative segment considered 30,202 hospitalizations.

**Results:** Our results confirmed the existence of strict small world structure in both intra- and postoperative surgical team networks. The U-shape trend of the small world property indicates a decreasing trend in the small world property after an increasing period over time. Therefore, a sudden declining trend is expected in the future in both intra- and postoperative networks, since the small world property is currently at its peak. In addition, high network density was observed in the intraoperative segment and partially in postoperative one, representing the existence of cohesive clusters of providers. We also observed that the small world property is exhibited more in the intraoperative compared to the postoperative network. Analyzing the temporal aspects of the networks revealed the postoperative segment tends to lose its cohesiveness as the time passes. Finally, we observed the small-world structure is negatively related to the patients' outcome in both intra- and postoperative networks whereas the relation between the outcome and network density is positive.

**Conclusion:** Small changes in graph-theoretic properties of the intra- and postoperative networks cause changes in the intensity of the structural properties. However, due to the special characteristics of the examined networks (*e.g.* high interconnectivity, team-oriented), the network is less likely to lose its structural properties unless the central hubs are removed. Our results highlight the importance of stability of personnel in key positions. This highlights the important role of the central players in the network that offers change-leaders the opportunity to quantify and target those nodes as mediators of process change.

## INTRODUCTION

Graph theory and network analysis have been widely used in mathematics and computer science to model pairwise relations between given objects and to analyze their interrelationships. For example, they have been used to simulate and study the spread of diseases [1], the evolution of the structure of the internet [2], or the study of genetic regulation in molecular biology [3] by focusing on connections among large numbers of features rather than the features themselves. Social network analysis has been recently used in health care settings as a technique for analyzing diffusion of new practices as well as collaboration and knowledge sharing among health care providers [4-6].

The structure of the network can play a significant role in determining the probabilistic behavior of the network. In the late 1950s, Paul Erdős and Alfred Rényi, introduced two random graphs models that changed the world of combinatorics. The Erdős-Rényi random graph is the original random graph model [7]. Erdős and Rényi showed that many properties of random graphs can be calculated analytically, and such network analyses can be applied to many real-world problems. However, the discovery of multiple types of prototypical structures in the late 1990s, as opposed to the universal random structure originally proposed by Erdős and Rényi in 1960 [7] facilitated a renaissance in the science of network topology.

Interactions among health care providers in a hospital not only can affect the culture but the knowledge level and efficiency of the entire system [4]. Additionally, maintaining collaborative behavior and supporting integration among providers would contribute to the multidisciplinary approach to health care [8], which is critical for patient safety and quality of care [9]. Moreover, mutual communication among the team members facilitates the information flow that may affect the patient outcome [10]. Network analysis can be employed to analyze such complex and sensitive networks in health care through modeling the interactions among the providers' teams and professionals [11], identifying the gaps and malfunctions that may affect patient safety, quality of care, and patient outcome [12].

Teams are realized as the core unit through which tasks are accomplished in organizations [13]. The dominance of team structures in organizations has been backed up by a strong theoretical and applied research [14]. Cohesive and well-functioning providers' team combinations contribute to providing monolithic high-quality care during hospitalizations [15], while considering the multidisciplinary nature of the teams in health care settings. Considering the size of the teams and number of providers involved, sustaining an effective teamwork and high performance is challenging [15]. The mechanism that is applied to team formation also affects the team network structure and the performance level [16]. In fact, the team function can be evaluated by social network analysis and the working processes can be verified and improved [17, 18].

The network structure may affect performance of the members [19, 20], however, studying complex networks is comprehensive task [21]. Several studies focused on modeling complex networks and analyzing the impact of macro- (*e.g.* clustering coefficient) and micro-level (*e.g.* motifs) features on the network structure [22, 23]. Small world is a network structure in which although most of the nodes are not directly adjacent, the nodes can be accessed through any node in the network by a few number of links [24]. In small world networks, although most nodes are not directly linked together, they can be reached by a small number of steps from every other node in the network. In addition, it is argued that the information is spread more efficiently in small world networks. This can facilitate the flow of knowledge [25], as well as team work. It is argued that even a network with significant amount of randomness can become small-world in certain settings [26]. More interestingly, once a network obtains the small world property, it will retain it with a high probability, even if the network experiences significant shocks [27] in terms of any type of disturbance or disruption to the network that might disconnect vital/important links among the nodes.

Another important network structure is called scale-free in which the degree of the nodes follows a power law distribution [28]. Such networks exhibit certain characteristics. One of the most interesting features of such networks is the existence of a few nodes with a high number of connections forming the core of the network. In scale-free networks, new nodes are more likely to connect to such highly connected nodes [29]. Thus, the core nodes are placed in the middle of network activities while being surrounded by progressively lower-degree nodes. This hierarchy in the nodes' connection structure makes the network fault resistant such that failures of a random node are more likely to influence a less-connected node versus one of the rarer, highly connected nodes. Therefore, such topology is useful in a highly sensitive structure that might need fault tolerance.

Our objective in this study was to examine the structure of intra- and postoperative case-collaboration networks among the surgical service providers in a quaternary-care academic medical center using retrospective electronic medical record (EMR) data. Case-collaboration describes situations where different healthcare providers worked together in the perioperative care of surgical patients, both within and across surgical, anesthesia, and nursing roles. Understanding the structure of the network will help to identify if the surgical service providers are collaborating systematically (*e.g.*, rather than randomly) in a well-established network, to determine the types of systems-level risk of node and edge failure as a function of network structure, and to offer additional hypotheses concerning how such collaborations are expected to evolve over time. Fig. 1 depicts two one-week snapshots of the surgical service providers' networks in intraoperative and postoperative phases. As seen, both networks are highly interconnected, forming a large core of providers in the center of the network. The postoperative network (Fig. 1-a) is bigger than the intraoperative one (Fig. 1-b), and as a result denser. The core in the postoperative network is surrounded by few dense clusters of providers. The hyper core team as well as the peripheral clusters indicate existence of implicit groupings. Same is observed in the intraoperative phase where providers are grouped in three big hyper cores. Based on the observed properties, we hypothesize the existence of a small world structure in the examined networks at the provider level[1], as the small world structure is expected to be observed in team-oriented networks [16, 30], and given prior simulations that anticipate the evolution of a small-world, scale-free network within an operating room environment [31].

---

[1] Providers' role-specific network structure analysis is out of the scope of this study as we wanted to analyze the structure of the intra- and postoperative networks, and their trends, at the global level. However, analyzing the network structure of different roles can be considered as a future research direction.

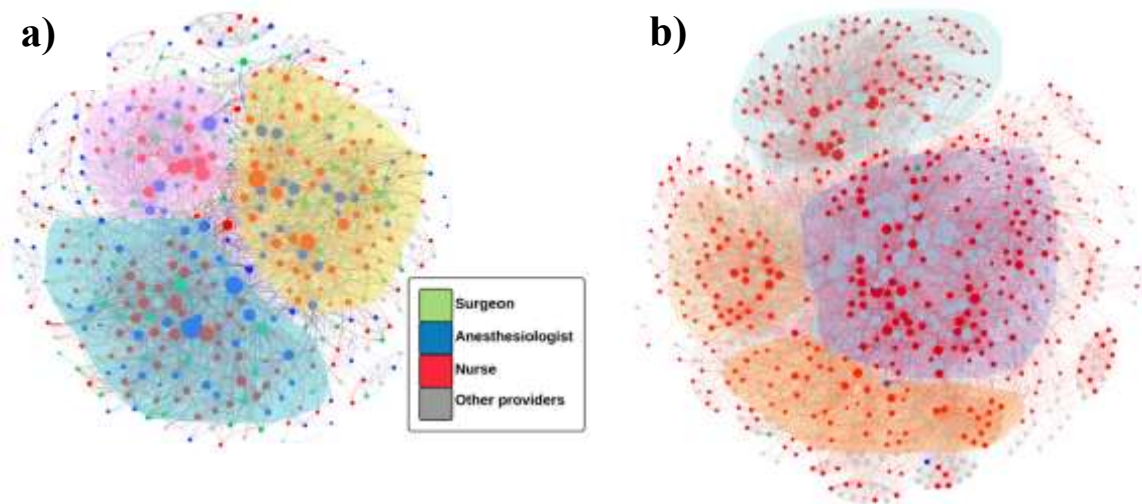

**Fig. 1** Sample role-specific one-week surgical service providers networks, **a)** Intraoperative, **b)** Postoperative. In the figure, surgeons are highlighted in ***green***, anesthesiologists in ***blue***, nurses in ***red***, and other providers in ***grey***. The colored regions indicate clusters of providers. Node sizes reflect number of connections (degree) of the node. Both networks are highly inter-connected. Nurses dominate the postoperative network. In postoperative network, a hyper core is observed which is surrounded by smaller size clusters of providers' teams, whereas in the intraoperative network three large clusters are observed.

## MATERIALS AND METHODS

The University of Florida Institutional Review Board (IRB) approved this study (IRB number 201400976). The data for this research were collected from the University of Florida's Integrated Data Repository (IDR) after obtaining a confidentiality agreement from the IDR.

### Data

The data contained de-identified intra- and postoperative information for adult patients, ages ≥ 21, who received nonambulatory/nonobstetric surgery at Shands at the University of Florida between June 1, 2011 and November 1, 2014, although the actual start date of data collection may have been delayed by the IDR by up to 90 days to aid in the de-identification of subjects. Subjects who did not receive a surgical procedure or who were discharged on the same day as their surgery were excluded from the dataset. We checked for possible duplicates and removed them from the data. The intraoperative segment contained 30,245 surgical cases being treated by 661 surgeons, 628 nurses, and 296 anesthesiologists. The postoperative segment considered 30,202 hospitalizations, 241 surgeons, 2,124 physicians, 3,088 nurses, and 102 anesthesiologists. The dataset also included a de-identified time order of the patients receiving treatment. Complications were based upon ICD9 codes captured during chart abstraction following the end of the hospital encounter; importantly, such codes simply denote an occurrence during the hospitalization and do not necessarily reflect causation, effect, or even association with a given surgical procedure or perioperative staff member.

### Networks, Basic Terminology and Definitions

Graph theory[2], *i.e.* the study of graphs, is traced back to the "seven bridges of Konigsberg" problem[3] of Leonard Euler in 1736 [33]. A graph is a symbolic representation of a network describing the inter-relations among the members of the network. Thus, a graph can be regarded as a set of (connected) nodes which reflect an abstraction of a real-life problem. Mathematically, a graph $G(n,e)$ is defined as a set of nodes $n$ and edges $e$. A node in graph $G$ is a terminal point or an intersection point, and an edge is a link between two given nodes.

---

[2] The word "graph" was first used by Sylvester in 1878 [32].
[3] In this problem, a man has to cross all the seven bridges once and continuously. Euler represented the problem as a set of nodes and edges and proved that the problem has no solution!

Graphs can be used to represent the inter-relations and processes in many real-life problems [34-37]. In a care setting, for example, medical service providers can be considered as nodes of the network and any type of relationship between them, such as being involved in a surgery, can be regarded as edges that connect the network nodes to each other. The formation of nodes and edges varies in different networks, leading to different network structures which exhibit a diverse set of properties. One of the common distinctions between different graphs is the way that edges connect the nodes. A graph is called an *undirected graph* if all the edges are bidirectional, and *directed graph* if edges are directed from one node to another.

**Network Variables Definition**

To investigate the small world property in the intra- and postoperative networks, we used the small world indicator that is calculated based on the clustering coefficient and the average path length in a network.

*Clustering Coefficient*

Clustering coefficient measures the level of the tendency of the nodes to cluster together [38], and is defined based on the number of triangles in a given undirected graph. An undirected graph is a graph in which the edges are bidirectional, and a triangle refers to any set of three nodes that are all connected to each other. Watts and Strogatz [24] define the local clustering coefficient for the node $i$ ($LCC_i$) as in Equation (1).

$$LCC_i = \frac{Number\ of\ triangles\ connected\ to\ node\ i}{Number\ of\ open\ triples\ centered\ on\ node\ i}. \qquad (1)$$

Here, the denominator counts the number of sets of two edges that are connected to the node $i$, i.e. open triples. Hence, a triangle is a closed triple of nodes where all three nodes are interconnected. The local clustering coefficient is used to calculate the overall clustering coefficient of the network ($CC$) as in Equation (2).

$$CC = \left( \sum_{i=1}^{n} LCC_i \right) / n. \qquad (2)$$

In Equation (2), $n$ is the number of nodes in the given network. The value of $CC$ ranges from 0 to 1, such that a larger number means higher interconnectivity of the network, while a smaller number indicates less interconnectivity. In other words, higher clustering coefficient can be reflected as more cliquish networks as the nodes in the graph tend to cluster together more.

*Shortest Path*

Shortest path length represents the minimum number of intermediary nodes that should be traversed from a source node to reach to a destination node in a network [39]. In the surgical service network, the shorter the distance is, the more easily the surgical service providers can get in contact with each other and can collaborate. In our study, this measure was calculated for the largest connected component of each network, and as a result, the small world indicator is also reported for the largest component. The largest component of a network is a subnetwork in which there is no isolated node and all the nodes are interconnected. The assumption of calculating the small world indicator for the largest component has been widely used in several studies [19, 20, 40, 41] and is easily justifiable as the core activities are expected to occur mainly in the largest component of each given network in which the most important and active nodes are present [42]. We measured the proportions of the largest component in our networks for both intra- and postoperative datasets. According to Fig. , the largest component covers nearly all nodes (>99%) in both examined datasets. In particular, except for the first and the tenth periods in the intraoperative dataset where the proportion of the largest component was ~99.8%, we obtained 100% coverage of the largest component in all the other periods in intraoperative as well as postoperative datasets. Thus, our assumption in calculating the path length in the largest component is strongly validated.

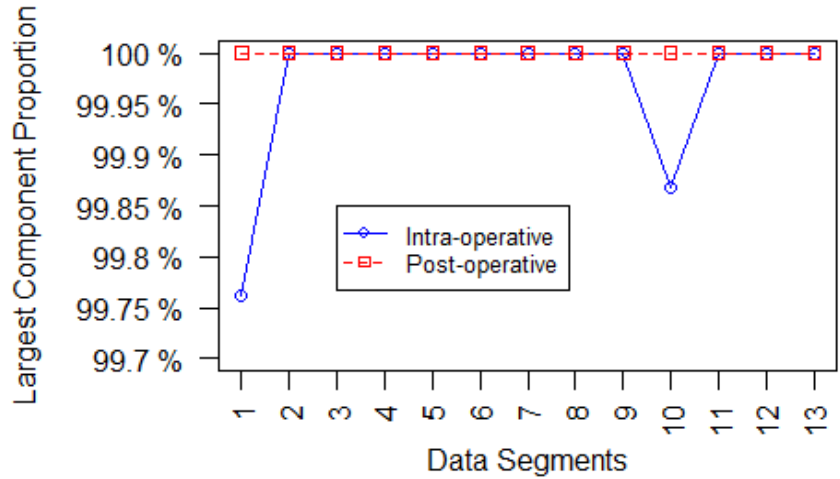

**Fig. 2** Proportion of the largest component in postoperative (red) and intraoperative (blue) segments. The largest component covers all the nodes for all the time events in the postoperative segment. The first and tenth intraoperative data segments are not completely covered by the largest component, but the proportion is highly significant, covering more than 99% of the nodes.

## Methodology
In this section, we describe the methodology of network construction and analysis in detail.

## Data Segmentation
Using the de-identified time order, we sliced the data into 13 separate, sequential time intervals, each containing 100 days, spanning from 0 to 1300. This step was necessary to explore the evolution of the network structure over time and to analyze the structure of the surgical team network more accurately. At the end of this stage, 13 different data slices were created for both intra- and postoperative datasets. The main reason for choosing a 100-day window was that we were interested in evaluating the seasonal impact, thus a ~90-day window was considered to further protect the identities of the subjects.

## Network Analysis
Surgical service providers can be considered as nodes in the medical collaboration networks and any type of relationship between them, such as being involved in the same surgical procedure, can be regarded as edges that connect the network nodes to each other. We constructed unweighted undirected networks for each of the 13 time segments separately for intra- and postoperative datasets. For this purpose, we first created the two-mode networks [39] of surgical service providers. Two-mode networks are called *bipartite graphs* in graph theory in which nodes can be divided into two disjoint sets and edges can only connect a node from one set to a node in another set, hence there is no inside-set connection. In our intraoperative 2-mode networks, surgical service providers, *i.e.* surgeons, anesthesiologists, and circulating nurses, were connected through the surgical cases performed by them. In other words, in our two-mode networks, all the surgical service providers who were present in (or in contact with) a patient's operation are connected to the surgical case (Fig. , left). For postoperative 2-mode networks, surgical service providers were connected through the hospitalization of patients. Because service providers were not connected to each other in the 2-mode networks, we converted each of the created 2-mode networks to 1-mode networks in which surgical service providers are connected to each other if they have been involved in the same surgical case in intraoperative segment (Fig. , right), or in the same hospitalization of a patient in postoperative segment. We focused on 1-mode networks since providers' networks and their interconnections were of our interest in this study.

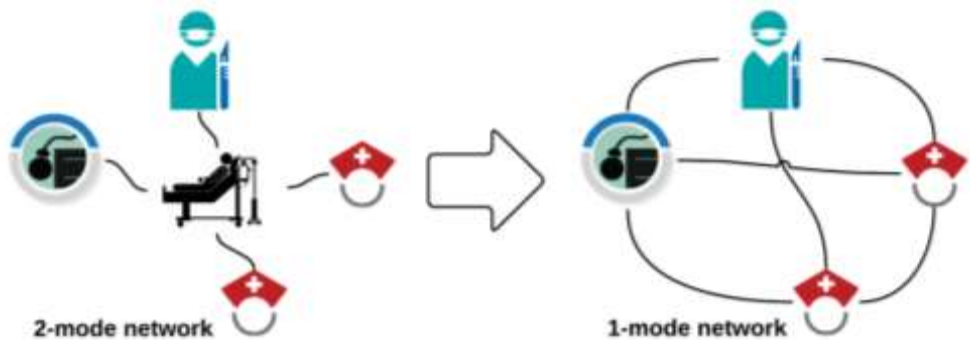

**Fig. 3** 2-mode vs. 1-mode networks. The 2-mode network is converted to a homologous 1-mode structure by removing the patients from the 1-mode networks. The edges thus denote a connection between healthcare providers via a shared patient encounter.

We then calculated the small world measures and analyzed the small-world property in each of the one-mode networks. Additionally, the trend of the small word property over time was also investigated. The entire procedure is presented in Fig. 4. According to Kogut and Walker [27], if a network has small world structure, it should satisfy two main conditions: 1) Its clustering coefficient should be significantly higher than a generated random network of the same size (same number of nodes), and 2) It should have approximately the same path length. This is because, by definition, small world networks exhibit a high clustering coefficient with a relatively short path length [43]. Therefore, to investigate the small world property in the collaboration network of surgical service providers, we compared the actual networks with two different reference models: 1) The Erdős–Rényi random network [11] with the same number of nodes and edges, and 2) The configuration model [44, 45] that allows to create networks with the same degree distribution as the actual networks. The comparison was made for each of the data segments in both intra- and postoperative datasets. In Erdős and Rényi's random graph [7], an edge is created between each set of two nodes independent of any other edges in the graph and with equal probability. In the configuration model, the network is generated using a prescribed degree distribution, in our case, the degree distribution of the actual network.

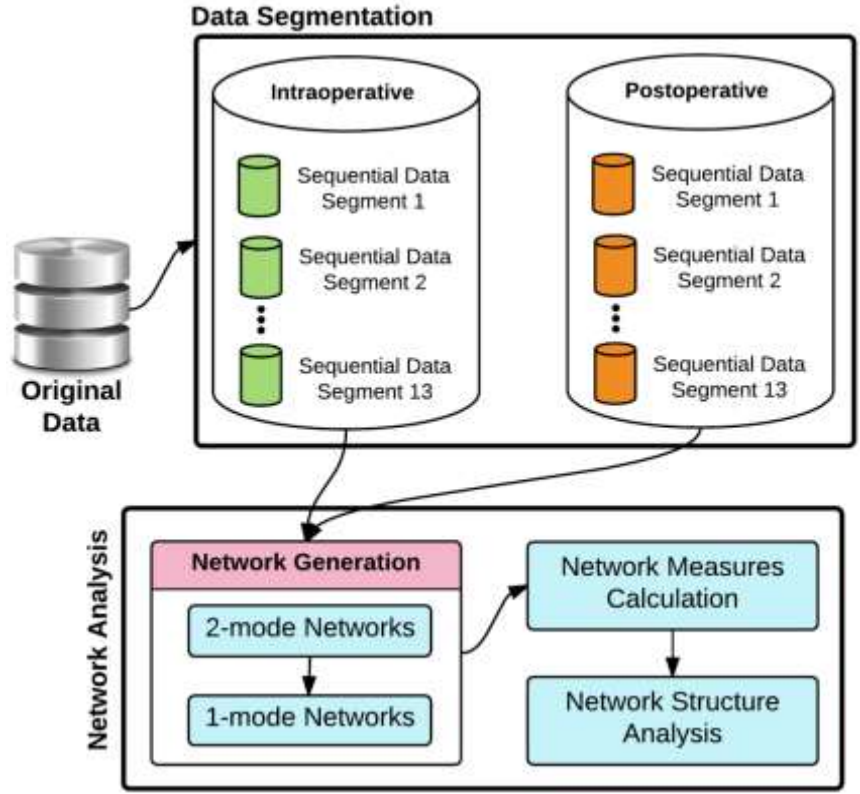

**Fig. 4** The original data for both intra- and postoperative datasets were first sliced into 13 separate, sequential data segments (smaller data disks). Next, we created surgical team networks for each of the 13 data segments for both intra- and postoperative datasets. We then calculated the network structure variables. In the final stage, we used the calculated network measures to analyze the structure of the examined datasets.

The small world indicator can be used to analyze the trend of the small world property over time. Having calculated the clustering coefficient (*CC*) and the path length (*PL*) as defined in Equations (1) and (2), the small world indicator is defined as in Equation (3).

$$SW = \frac{\frac{CC_a}{CC_{rnd}}}{\frac{PL_a}{PL_{rnd}}}. \qquad (3)$$

In Equation (3), $CC_{rnd}$ and $PL_{rnd}$ are the clustering coefficient and path length in the random networks, respectively. This approach for calculating the small world indicator has been applied in several studies across multiple disciplines [19, 27, 46]. We used the same formula for calculating the small world indicator with the configuration model as the reference network. Using the small world indicator calculated for each of the segments, we analyzed the small world property trend in surgical service networks over the entire time interval.

Using the calculated network measures, we also did a cohesion analysis on the intra- and postoperative networks. Cohesion is defined as relatively dense and highly connected subgroups in a network where members are extensively and very frequently in contact, and relate easier to the members of their own subgroup than the ones of other subgroups [47]. It reflects the degree the nodes are connected to each other, their reachability and closeness. In highly cohesive networks paths among the nodes are relatively shorter, thus the nodes are closer [48]. We analyzed four measures representing the network cohesion: 1) The ratio of the number of triangles, *i.e.* number of three nodes that are all adjacent, to all possible triangles in the network, 2) The average shortest path length, 3) Network density, *i.e.* the ratio of actual connections to all possible connections, and 4) The structural holes [49, 50]. Structural holes are based on the theory that nodes in a network possess specific positional advantages and drawbacks and the structural holes is defined as the gap between two nodes in a network that have complementary resources/information/etc. We used Burt's measure of constraint to quantify the structural holes concept:

$$n_{ij} = \frac{w_{ij}}{\sum_k w_{ik}}. \quad (4)$$

In Equation (4), $w_{ij}$ is the values of the edge from $i$ to $j$, *i.e.* the number of times the surgical service providers $i$ and $j$ operated a patient in the same surgery[4], and $n_{ij}$ calculates the ratio of $i$'s connections with $j$ compared to all connections of $i$. From $n_{ij}$ a network containing dyadic constrains ($c_{ij}$) is generated as follows:

$$c_{ij} = \left( n_{ij} + \sum_{k, k \neq i, k \neq j} n_{ik} n_{kj} \right)^2. \quad (5)$$

The aggregate constraint vector ($C_i$) is calculated by summing up the $c_{ij}$ values over $j$. We calculated the aggregate constraint vectors for each of the intra- and post-operative networks and considered the average of the vector values as the structural hole measure of each network.

Finally, we performed a correlation analysis, separately for intra- and postoperative networks, to assess the relationship between the small-world property and network cohesion with patients' outcome, represented by number of complications. We did a robust correlation analysis to address the anomalous data and prevent biased results due to the existence of outliers. A projection-based estimation procedure, *i.e. Stahel-Donoho* method [51-53], was used to estimate the sample mean and covariance matrix. In other words, this method identifies outliers through viewing the data from the right perspective. The robust distances are first calculated via a projection computation, and then they are used in a weight function to calculate the weighted mean and the covariance matrix [54]. We used the *CovRobust* function in the *rrcov* library in R programming language [55] which provides robust estimates of the location and covariance.

## RESULTS

We first focused on the size of the networks in intra- and postoperative datasets and analyzed their trends in all time slices. As seen in Fig. 5 the size of the postoperative network is, on average, nearly threefold larger than the intraoperative network within each data segment. In addition, in the period of 1201-1300 days, *i.e.* the last data segment, both the intra- and postoperative segments contain fewer observations, which were expected, as the final slice is smaller in both datasets (covering ~60 temporal events rather than 100). Fig. 5 also highlights that the distribution of data is comparable over the time slices for each of the examined datasets. This permits comparison between the resulting network and the respective findings of each time slice to one another. Finally, the stable trend for both datasets partially confirms the equality between the number of inputs (newcomers) and outputs of the system during the observed period. This is important given that the institution in question is an academic training facility, with annual admission and graduation of interns, residents, and fellows. Note that in general, the size and duration of these training programs are largely driven by requirements from the Accreditation Council for Graduate Medical Education, thus minimizing large changes in the structure of training programs from one year to the next.

---

[4] Since we constructed undirected unweighted networks of the surgical service providers, $\forall i, j$; $w_{ij}$=1.

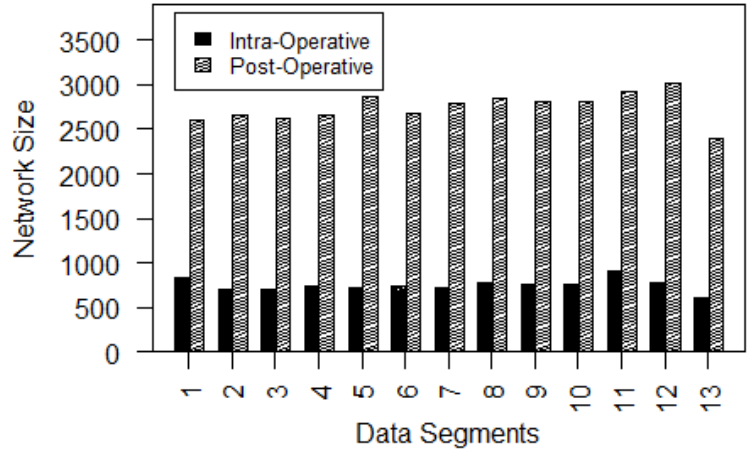

**Fig. 5** Network size in postoperative (shaded) and intraoperative (solid) data segments.

## Small World Network Analysis

We tested for the presence of the small world property and evaluated its trends in the intra- and postoperative segments. We used two reference models for comparing the clustering coefficient and path length of the actual networks: 1) random networks, 2) configuration network. As seen in Fig. , the clustering coefficient in each of the segments in the intra- and postoperative segments is significantly higher than the respective generated random networks. This represents the high tendency of teamwork among the providers in a given surgical service team. The examined networks are an excellent example of a tight and dense interaction among the nodes, which is a primary sign of the existence of the small world structure.

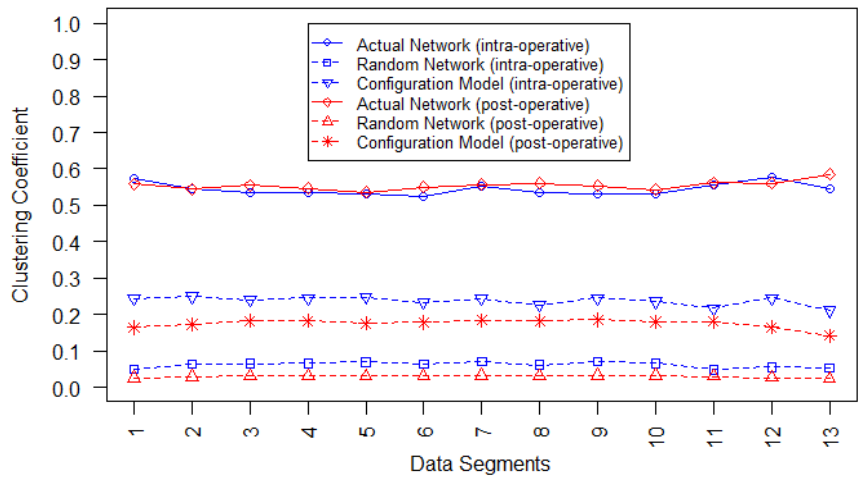

**Fig. 6** Clustering coefficient over time in postoperative (red) and intraoperative (blue) segments. Solid lines represent the clustering coefficient in actual networks and dashed lines show the values in random networks. Clustering coefficients in both segments are much higher than the values in respective random networks.

Next, we compared the path lengths in actual and random networks (Fig. 7). Interestingly, path lengths of the actual and generated random networks, which are constrained to the same size during generation, almost match. This is even more evident in the intraoperative dataset (Fig. 7, blue lines) where the path lengths are almost the same. For the postoperative data, the path lengths are very close in the final period. According to the Watts and Strogatz [24] definition and based on the findings from figures 6 and 7, both intra- and postoperative networks clearly and strictly resemble the small world property.

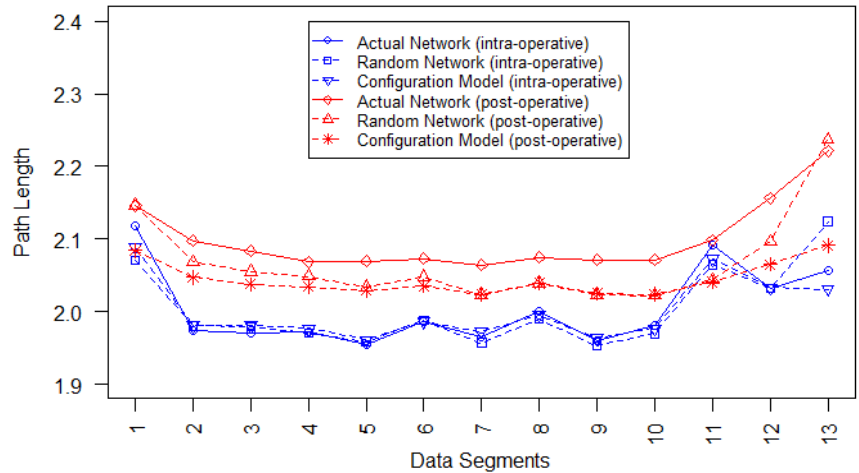

**Fig. 7** Path length over time in postoperative (red) and intraoperative (blue) segments. Solid lines represent the path lengths in actual networks and dashed lines show the values in random networks. Path lengths in both segments are very close to the values in respective random networks.

We also analyzed the small world property [as defined in Equation (3)] trend over the examined sequential data segments Fig. 8). A steady trend in the small world indicator is observed within the intermediary periods, before increasing drastically during the final periods. This is in line with the nature of the small world structure. As new providers join the systems, the system is not at first using all of its potential for creating new teams. As time passes, more co-working links (mostly new pairs) are established that might cause the network to evolve dynamically and reflect more of a small world property. Another observation is the U-shape trend of the small world indicator in both examined datasets, which is again in agreement with the literature [56]. This U-shaped trend indicates a decrease in the small world property after an increasing period. As the small world property is almost at its maximum in both examined datasets, a sudden declining trend is expected in the future. The reason is that the small world network collaborators gradually get connected to the brokers (nodes able to connect different clusters), who enable them to have access to other clusters and communities. This provides them with a diverse set of coworkers that might help them to maintain or even improve their position in the network [57]. The U shape of the small world indicator also shows that as the time passes, new(er) nodes gradually get familiar with the important (old) brokers as well as denser clusters, resulting in more homogeneous clusters. Therefore, the level of diversity in the network decreases and the position of the old brokers also decays. This might result in a higher rate of intra-cluster collaboration among the already known partners, which will lead to a less small world property [56].

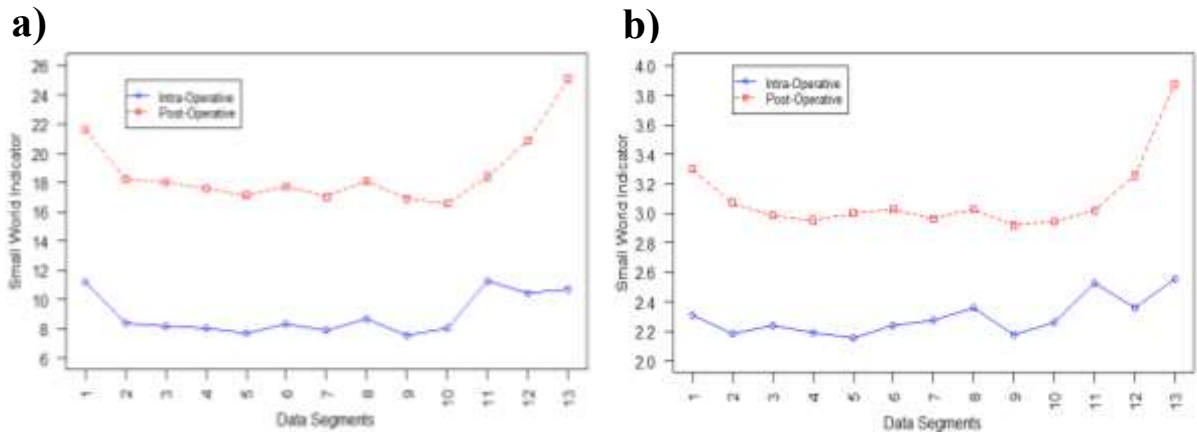

**Fig. 8** Small world indicator in postoperative (dashed line) and intraoperative (solid line) segments, **a)** using random networks, **b)** using the configuration model. The U-shape of the small world indicator is observed in both segments. Although the small world indicator would be greater than 1 in small world networks, one may note that having a small world indicator greater than 1 is not sufficient to prove the small world property in a network, and both $CC_a/CC_{rnd} > 1$ and $PL_a/PL_{rnd} \sim 1$ are required.

From Fig. -a and b, it seems that the postoperative dataset exhibits the small world property more due to a higher level of the small world indicator. But, it is the intraoperative network that shows a higher degree of small worldness! The small world indicator is useful for analyzing the evolution of the small world property within a dataset, but is not appropriate for comparing the level of small worldness in intra- and postoperative datasets, as it is dependent on the size of the network [58]. As previously discussed, the clustering coefficients for the examined networks are comparable (Fig. ). However, the path length values in the intraoperative networks are very close to the ones in the respective random networks (Fig. ), thus intraoperative network shows small world property more. Interestingly, using random networks and configuration model as the reference network resulted in almost the same trend for the small world indicator, with a lower range observed for the configuration model. However, the small world property requirements are satisfied in both approaches, confirming the existence of small world structure in the surgical service providers networks.

**Scale-Free Network Analysis**

We showed that the small world property is strictly exhibited by both intra- and postoperative segments. In another study, we explored the scale-free property in these networks [59]. In scale-free networks, some nodes are expected to have degree higher than the average degree. Our findings confirmed that in both intra- and postoperative segments, there are enough candidates with high degrees. Additionally, we found that the intraoperative segment is more likely to be a scale-free network rather than the postoperative one. Finally, it was observed that as time passes, the postoperative segment loses its scale-free property, which can be mainly due to the consequences of the developmental mechanisms through which the connections among the surgical service providers are formed.[5]

**Cohesion Analysis**

It was observed that the surgical service providers networks are highly interconnected, exhibiting both small-world and scale free properties. In these networks, there are highly central providers, named hubs, who act as brokers, connect separate clusters of providers, and attract more connections. Humans tend to form cohesive clusters [60]. As part of the small-world analysis, we found that the examined networks have an average shortest path of ~2, meaning the nodes are on average accessible through only one intermediary node. The low observed shortest path can be also interpreted as a sign for cohesive networks as members in a highly cohesive network are reachable easier as they are connected through less number of links. We further analyzed network cohesion by focusing on three other aspects of the networks over time, 1) The ratio of the number of triangles, *i.e.* a set of three nodes that are all adjacent, to all possible triangles in the network, 2) Network density, *i.e.* the ratio of actual connections to all possible connections, and 3) The structural holes.

---

[5] For more details on the scale-free network analysis, please refer to our paper on scale-free characteristics of surgical networks [59].

As seen in Fig. 9, the intra-operative networks have on average more triangles. As the proportion of triangles can represent small cohesive sub-groups in the network, in the intra-operative networks we observe more small-size cohesive interactions. From figures 6 and 7, it is observed that the number of triangles follows an opposite direction compared with the small-world indicator. This may indicate the importance of dense interactions among relatively bigger teams to enhance the small-world property. Same observation, with lower intensity, is also made for the post-operative networks.

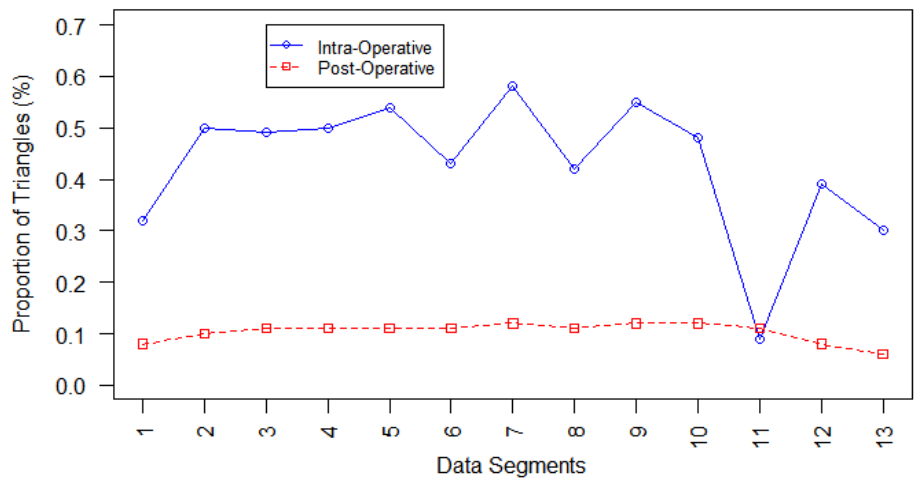

**Fig. 9** Percentage of number of triangles to all possible triangles in the network. The intra-operative networks have more triangles compared to the post-operative networks. More triangles can be regarded as a sign of small-size cohesive networks.

The analysis of the number of connections to all possible connections, *i.e.* network density, also complies with the analysis of the number of triangles. Fig. 10 Shows the results. Therefore, one reason for having more small-size subgroups in the intra-operative networks could be higher density which increases the possibility of forming triangles among the surgical service providers. Again, the trends of intra- and post-operative network density follows the small-world indicator with an opposite direction, hence the previous argument is still valid.

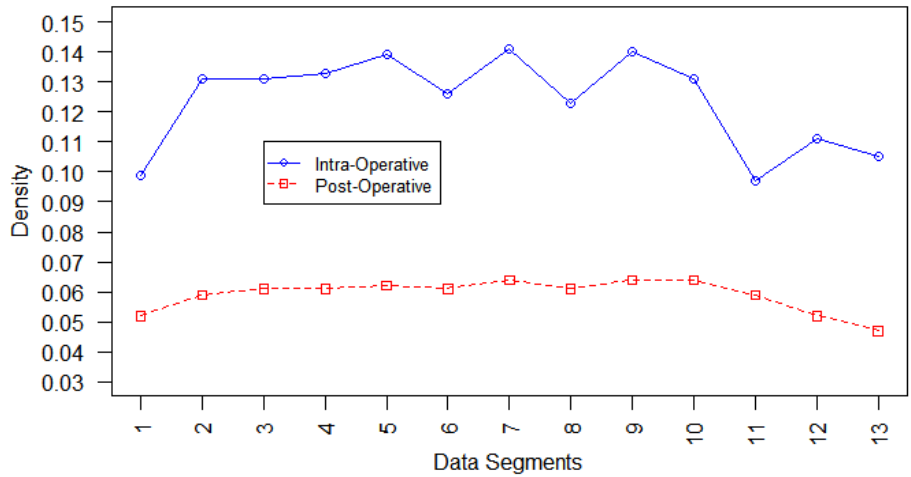

**Fig. 10** Network density in intra- and post-operative networks. Intra-operative networks that are smaller in size are denser than post-operative networks.

Fig. 11 Shows the results for the structural holes analysis calculated by the aggregate constraint measure. In general, lower constraint value indicates more structural holes that can be exploited by the members in the network to play the brokerage role. On the other hand, higher constraint values restrict the brokers in the network, making brokerage roles more critical as if brokers are diminished, the connections between different network clusters would decrease. As seen, the overall aggregate constraint in the intra-operative network is slightly higher than the post-operative one. The trend of the overall aggregate constraint follows the trend of the small-world indicator (figures 6 and 9). From Fig. 6-a, surgeons have higher constraint representing more opportunity for brokers to bridge among them and fill the structural gaps. In the post-operative network, different roles have higher structural hole risk in various data segments

**a)**

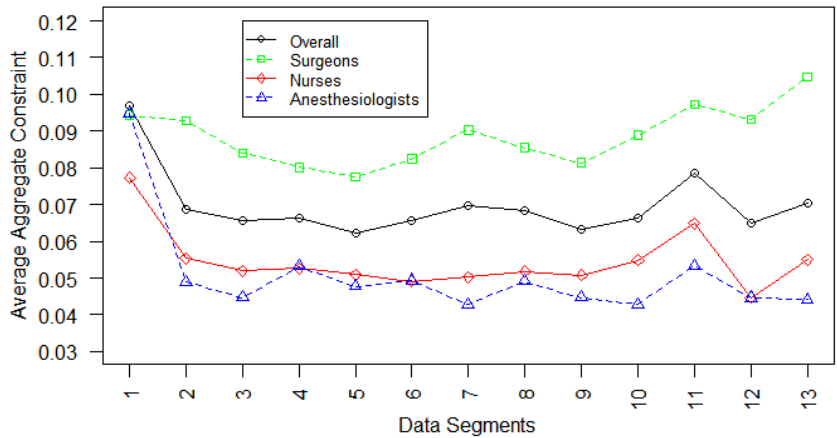

**b)**

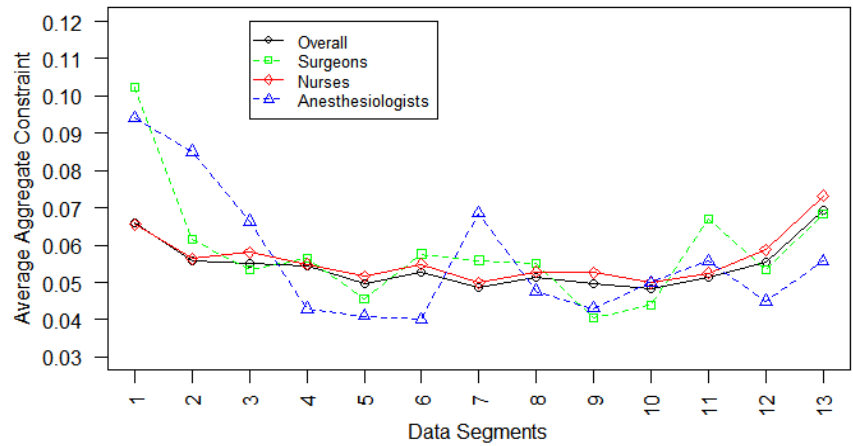

**Fig. 11** Average aggregate constraint (structural holes) in **a)** intra-, and **b)** post-operative networks. The overall aggregate constraint (the black line) follows the small-world indicator trend and is slightly higher in the intra-operative networks.

**Network Properties vs. Patient Outcome, Correlation Analysis**

We finally analyzed the linear relationship between the observed network properties, represented by the small-world indicator and network density, and patients' outcome. We performed a robust correlation analysis separately on intra- and postoperative networks to assess the relationships between the structural properties and patients' outcome. We considered the ratio of the number of complications to the number of surgical cases as a measure of patients' outcome, named as average number of complications. The results are shown in Fig. 12.

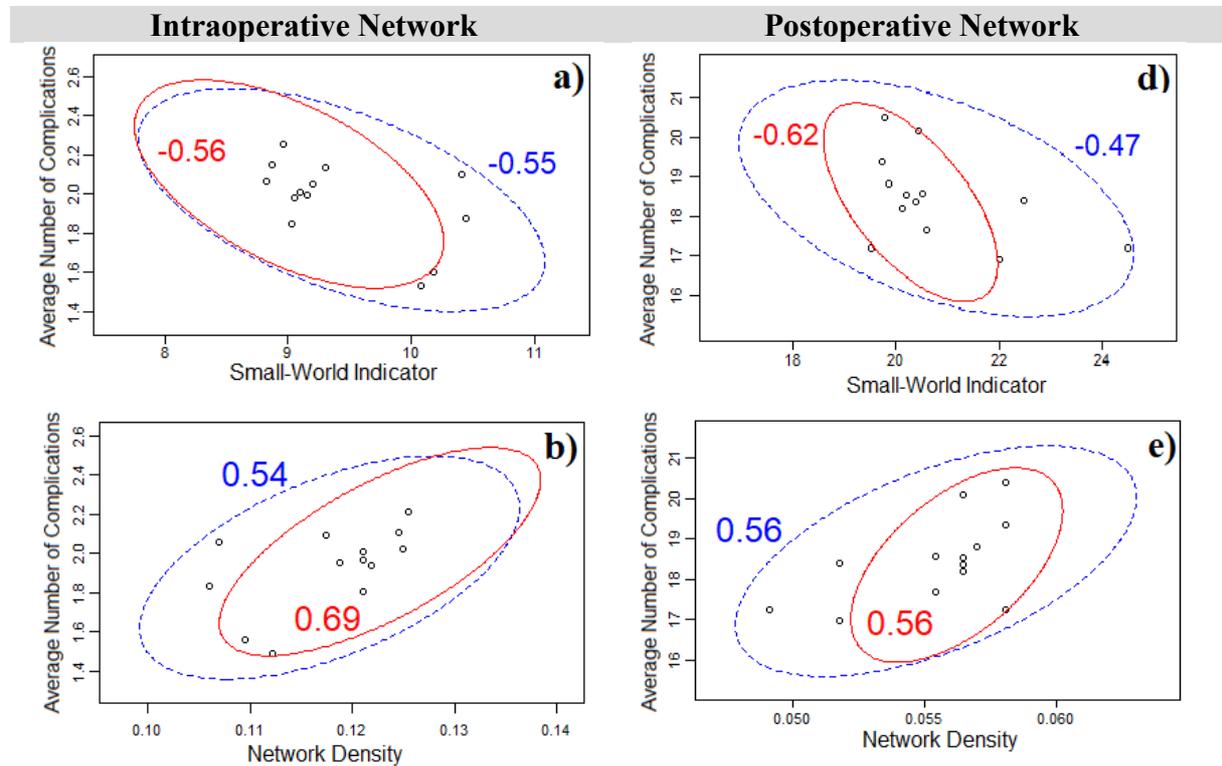

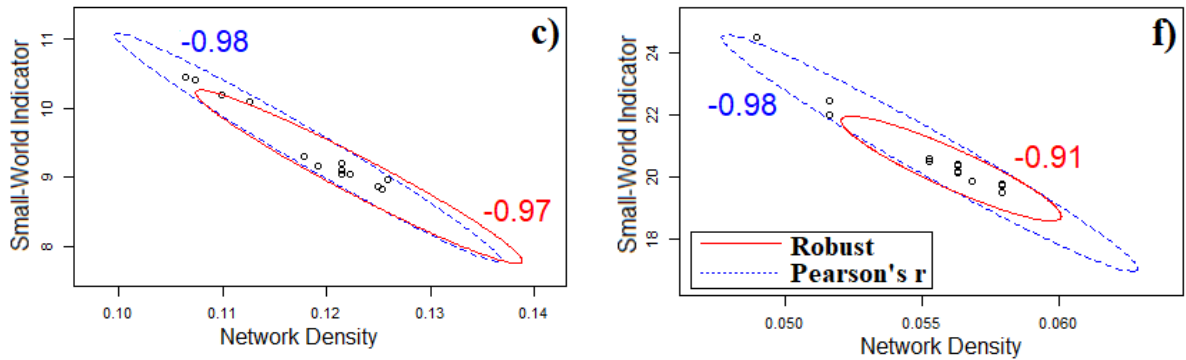

**Fig. 12** Pearson's r and robust correlation analysis in **a-c)** intra- and **d-f)** postoperative networks. All values are significant at 95% confidence interval. The red solid line indicates the data points involved in the robust correlation analysis and the clue dashed line shows the data points in classical Pearson's r. Red and blue values in the figures are the correlation coefficient for robust and Pearson's r respectively.

As seen, the small-world indicator is negatively related to the patients' outcome in both intra- and postoperative networks whereas the relation between the outcome and network density is positive. The high network density may result in many highly-cohesive subgroups inside the network with high interaction rate among the members. This may affect the network performance negatively through redundancy of the connections that may lead to higher number of complications. Alternatively, the high interaction rate may reflect the additional numbers of multidisciplinary specialists needed for patients presenting with more complex medical and surgical needs. On the other hand, the small-world property may ease the flow of information and enhance the accessibility in the network that may positively impact the network performance through decreasing the number of complications. As expected a negative relation was observed between network density and small-world property, as when the network becomes highly dense most nodes will be actual neighbors of each other contradicting the small-world property. One should note that this is a preliminary analysis and several other factors can play a role in affecting the patients' outcome.

**CONCLUSION**

We used two large datasets to analyze the network structure of intra- and postoperative data segments. The analysis revealed seven specific characteristics of the examined networks: 1) The proportion of the largest component was extremely significant, which allowed us to do a more accurate analysis, especially in calculating the path lengths; 2) High clustering coefficients in both networks compared to the same size random networks during the examined time period; 3) Almost comparable path lengths in both networks with the ones in the generated random networks; 4) Strict small world property in both networks; 5) Scale-free property in the intraoperative data segment as well as in most networks of the postoperative segment; 6) It seems that the postoperative segment is losing its scale-free property over time [59], and 7) The intra-operative network is denser than the post-operative one and possesses more small-size cohesive subgroups.

As the small world property [59] and cohesive structure were observed in most of the examined networks, we can say that surgical service providers tend to form internal groups in which everyone is directly/indirectly connected to everyone. Such groups can be referred to as *communities* that are complete graphs[6] in the ideal form. The members of a community might have connections outside of the community. However, there exist a few members who possess strategic network positions by being connected to many communities. These hubs are also responsible for maintaining the small world property. This fact is more likely to be true in the intraoperative segment as the proportion of triangles and network density is higher compared with the postoperative networks. Considering the fact that the examined networks also possess scale-free property, since the number of hubs is limited in a network and the majority of the nodes are of small degree, the likelihood that the network loses its cohesive structure is relatively small unless we take out all the major hubs at once. Therefore, such major hubs can be regarded as both the strength and weakness points of the network [61].

---

[6] A complete graph is an undirected graph in which every pair of distinct nodes is connected.

This carries important implications for the perioperative environment. Typically, the day-to-day management of a suite of operating rooms is "run" by a charge nurse and an anesthesiologist. These individuals are responsible for assigning staff to operating rooms and coordinating the often-dynamic staffing and scheduling requirements as surgical cases progress throughout the day. For instance, a sudden exsanguination, intraoperative code, difficult airway, or emergent case posting can swiftly alter planned resource allocations throughout the operating theatre. The charge nurse and anesthesiologist, in our experience, are incredibly busy individuals who may receive phone calls every minute or two for hours on end. A random network failure is less likely to affect one of the charge individuals compared with a less well-connected node in the operating room network. That said, the implications from our analyses suggest that a chance disconnection of a change nurse or anesthesiologist could wreak disproportionate havoc upon the structural functioning of this network.

By virtue of the preferential attachment property, it would be suggested that the most highly-connected nurses and anesthesiologists are likely to become busier, more rapidly, over time compared with a randomly chosen node within the network. In deciding whether a "busy day" for a highly-connected charge nurse/anesthesiologist represents a cyclical peak versus a trend, the structure of the operating room network would suggest that these roles may lean toward a steadily increasing trend. This may help operating room administration plan for future growth, for example, by developing local charges that offset some of the communication traffic and serve as local nodes to promote and facilitate growth within certain domains. Cohesion analysis can also shed light on the interactions among the team members in the surgical networks. It is expected that similar providers interact and collaborate more easily which may affect the overall performance of the network. Meanwhile, working in teams affects the individual members through the positive interactions as they become or perceive themselves more similar.

The role-specific structural holes analysis revealed the importance of the brokers in the examined networks. We observed surgeons to have the highest constraint values indicating the risk of surgeons' brokerage roles as if the main brokers are diminished (*e.g.* retired) the connection level between different network clusters would decrease that may result in less cohesive structure. Moreover, in general, it was observed that the postoperative networks have higher structural hole risk that calls for more attention in this regard. Having observed that the network structures are highly correlated with the patient outcome, maintaining the network properties over time would be critical as well. The small-world property, in particular, eases the flow of information and enhances accessibility in the network that may positively impact the network performance.

These findings are especially notable within the context of national policy discussions surrounding resident fatigue. In recognition of the detrimental effects of extended work hours of resident physicians upon both patient safety and resident well-being, the past decade has seen a concerted effort to reduce resident work hours. The decrease in work hours inevitably lead to an increase in transfer-of-care events that allowed for 'leaks' of critical information in communications from one healthcare provider to another. These structural characteristics may be important considerations in enhancing patient safety given the critical role of interpersonal communications in transfer-of-care events [62].

Importantly, our results point to a wide range of questions concerning the underlying clinical processes driving these findings. For instance, due to the deidentified nature of our dataset, it remains unclear which characteristics of central actors may drive their particular centralities. Central actors could reflect high-volume attending physicians, busy resident physicians assigned to a wide array of surgical attending physicians operating at all hours of night and day, or junior staff who have yet to identify their particular niche role within the perioperative theatre. Sub-specialization may also drive, or limit, actor centrality, and also raises the question as to how best to isolate the effects of subspecialty group effects from those effects of subspecialty members, certain features of which may indeed help define the subspecialty-ness of said group effects. What's more, while social networks tend to be considered as static networks (and, temporally, as a series of static networks), many of those involved in process improvement may instead be primarily interested in longitudinal network architectures that focus on day-to-day, or even case-to-case, networks rather than aggregate networks. These, and many more, opportunities each invite additional methodologic and application-based explorations in follow-up work.

Network analysis of perioperative teams offers many exciting opportunities to explore how to better care for patients. One example of how healthcare social network knowledge may influence intra-unit workflow patterns is the work of Creswisk *et al.* [10]. Here, the team used social network surveys to identify advice-seeking networks amongst physicians, nurses, and administrators to identify actual versus organizationally-mediated, theoretical networks of communication. Another example from our team [63] used social network analysis of an acute pain service to quantify the extent of consultation sources, as well as to identify those lines of communication with low redundancy and/or dependent upon very junior trainees. Regardless of the proposed improvement, such changes must be disseminated among the staff in an effective manner to assure uptake and maintenance of process changes. Network analyses, such as the one demonstrated here, offer change-leaders the opportunity to quantify and target those nodes most likely to be well-connected as mediators of process change.

## LIMITATIONS AND FUTURE WORK

The main limitation of this study was the way we measured the interactions among the surgical service providers. We were unable to catch the informal relationships among the nodes (*e.g.* friendship) that might have an impact on the team arrangement procedure. This type of data is never recorded but has an impact on the network structure and its performance. However, we believe that such EMR-based measures of networks may offer an important contribution to defining interaction networks given the size and quality of the available data. Moving forward, it will be important to integrate EMR-level measures of networks with specific, objective measures of interaction such as via video, proximity measures via smartphones, and even social media/text-based communication volume and timings to create composite measures of network performance. An important limitation concerning encoded complications related to the imprecise mechanisms in common EHR structures for aggregating ICD-based complication codes across hospital encounters, such that temporal and activity-sequence linkages are lost. Our results are thus only able to highlight associations between network structures and such complications; while many mediators of the association are plausible, addressing such relationships will require more advanced methodologies that examine graph subsets at greater detail and, likely, smaller scope. Moreover, there is still much work to be done to determine how best to integrate network-level information into formal machine learning approaches to forecasting clinical outcomes. We focused on the entire network of the surgical service providers, mainly because of the considerable proportion of the largest component that helped us to calculate the related network measures more accurately. However, another future direction might be the analysis of the network structure and its evolution at the role level of the surgical service providers, *e.g.* surgeon, anesthesiologist, and circulating nurse. Lastly, we constructed unweighted networks of surgical service providers as having the same providers operate the same patient multiple times was a rare event in our dataset. Of course, weighted networks could be another future research direction.

**Declarations**

*Ethics*

The University of Florida Institutional Review Board (IRB) approved this study (IRB number 201400976). The data for this research were collected from the University of Florida's Integrated Data Repository (IDR) after obtaining a confidentiality agreement from the IDR.

*Data Availability*

Unfortunately, we cannot share the data in a public forum/repository due to protected health information (PHI) concerns and patient privacy reasons. However, in the interests of transparency, we would invite any parties interested in collaborating with us to develop a business data use agreement that we would process in a standard fashion through the University of Florida, and which would enable sharing of data according to the existing processes.

*Authors' Contributions*

Conceiving and designing the experiments: AE, PJT, PR

Performing the experiments: AE

Analyzing the data: AE

Data/materials: PJT, LZ

Writing of the manuscript: AE, PJT, LZ, PR